\documentclass[preprint]{aastex}

\slugcomment{Draft for the Astrophysical Journal, \today}

\begin{document}

\title{Rotation of Horizontal Branch Stars in Globular Clusters}
\author{Alison Sills, M. H. Pinsonneault}
\affil{Department of Astronomy, The Ohio State University, 140 W. 18th
Ave., Columbus, OH, 43210, USA}

\begin{abstract}

The rotation of horizontal branch stars places important constraints
on angular momentum evolution in evolved stars and therefore
rotational mixing on the giant branch.  Prompted by new observations
of rotation rates of horizontal branch stars, we calculate simple
models for the angular momentum evolution of a globular cluster giant
star from the base of the giant branch to the star's appearance on the
horizontal branch. We include mass loss, and infer the accompanied
loss of angular momentum for each of four assumptions about the
internal angular momentum profile. These models are compared to
observations of horizontal branch rotation rates in M13.  We find that
rapid rotation on the horizontal branch can be reconciled with slow
solid body main sequence rotation if giant branch stars have
differential rotation in their convective envelopes and a rapidly
rotating core, which is then followed by a redistribution of angular
momentum on the horizontal branch. We discuss the physical reasons why
these very different properties relative to the solar case may exist
in giants.  Rapid rotation in the core of the main sequence precursors
of the rapidly rotating horizontal branch star, or an angular momentum
source on the giant branch is required for all cases if the rotational
velocity of turnoff stars is less than 4 km s$^{-1}$.  We suggest that
the observed range in rotation rates on the horizontal branch is
caused by internal angular momentum redistribution which occurs on a
timescale comparable to the evolution of the stars on the horizontal
branch. The apparent lack of rapid horizontal branch rotators hotter
than 12 000 K in M13 could be a consequence of gravitational settling,
which inhibits internal angular momentum transport. Alternative
explanations and observational tests are discussed.

\end{abstract}


\section{INTRODUCTION}

There is compelling evidence for extensive mixing in the envelopes of
low mass evolved stars which is not predicted to occur in classical
stellar models.  Rotation is frequently invoked, at least implicitly,
as the underlying agent which is responsible.  There has been
extensive work on rotational mixing in both low and high mass main
sequence stars \citep{MC91,TZMM97}, and phenomenological work on mixing
in evolved stars has recently been undertaken by several groups
\citep{C95,S97}. However, detailed physical models of giant branch
mixing have proven to be a significant challenge to theorists.
  
The largest uncertainty, in our view, has been the lack of constraints
on the angular momentum evolution from the main sequence to the first
ascent giant branch, the horizontal branch, and beyond.  In this paper
we examine the implications of measured surface rotation rates of
horizontal branch stars for angular momentum evolution on the giant
branch.  We will show that the combination of rapid horizontal branch
rotation and slow main sequence rotation places strong constraints on
the angular momentum evolution of giants.

The pioneering work of \cite{SM79} remains the single best physical
analysis of rotational mixing in evolved stars.  They investigated the
link between classical meridional circulation and the CNO anomalies in
giants, and stressed the initial angular momentum budget and the
rotation law in the convection zone. They concluded that meridional
circulation was generally consistent with the observational data,
provided that the rotation rate on the giant branch was sufficiently
high. The necessary rotation rates on the giant branch require a
rapidly rotating core on the main sequence, and the convective
envelope of a giant branch star cannot be rotating as a solid body,
since the required main sequence rotation rates would be much higher
than observed.  This study neglected the mixing of elements caused by
differential rotation with depth in the star, and did not include the
effects of mass loss.

Ideally, we would like to study the rotation rates of giant branch
stars directly. These stars show the strongest evidence for
non-standard mixing, and their evolution from main sequence stars is
direct and well understood. By looking at the rotation rates of giants
at different luminosities on the giant branch, we should be able to
determine their initial angular momentum profile as angular momentum
is dredged up by the deepening convection zone. We should also be able
to test the predicted correlations between different internal rotation
velocities and the observed surface abundances. However, since giant
branch stars are so large, their surface rotation rates are predicted
to be very small, much less than 1 km s$^{-1}$. The spectral resolution
required to observe such velocities is far beyond what can be done
today, although gravitational microlensing may make this possible
\citep{G97}.

Fortunately, horizontal branch stars have observable rotation rates,
and they provide insight into the interiors of giant stars. Since
stars lose mass on the giant branch, the surface of the horizontal
branch star was once inside the giant branch star. Stars at different
effective temperatures on the horizontal branch have lost different
amount of mass, and therefore can be used to test the angular momentum
distribution within giants. The rotation of stars on the horizontal
branch therefore provides an indirect test of the internal rotation of
the same stars in previous evolutionary stages.

Observations of rotation rates of horizontal branch stars are
scarce. Rotational velocities for horizontal branch stars with
temperatures between 7000 K and 11 000 K have been measured in 6
globular clusters and the field
\citep{P83,PTC83,P85a,P85b,PRC95,CM97}, and rotation rates for stars
up to 20 000 K have only been measured in M13 \citep{behr99}. The
cooler horizontal branch stars show a range of rotation rates, both
between and within clusters. These stars are rotating at between 10
and 40 km s$^{-1}$, much faster than the few km s$^{-1}$ rotation
rates seen for the Sun and other old main sequence stars. The usual
explanation for this increase in rotation rate between the main
sequence and the horizontal branch is that the cores of main sequence
stars retain a substantial fraction of their initial angular momentum
and are rotating rapidly. This scenario, which conflicts with
helioseismic observations of the slowly rotating solar core, will be
tested in this work. 

Almost a decade ago, \cite{PDD} studied the angular momentum evolution
of stars as they evolve up the giant branch on to the horizontal
branch as a constraint on the amount of internal rotation
allowed. They found that differential rotation with depth, rather than
solid body rotation, was consistent with the existing observational
data. Since that paper, there have been a number of important advances
which have prompted us to follow up their work in this paper. Firstly,
the helioseismic observations of the rotational splittings of solar
p-modes strongly suggest that the Sun is rotating almost as a solid
body down to 0.2 R$_{\sun}$ \citep{Chap99,Laz96,Cor98}. Secondly, a
stringent test of halo star rotation comes from very high resolution
and signal-to-noise studies of the $^6$Li to $^7$Li abundance ratio in
metal poor halo stars.  No evidence for extra line broadening above
the 4 km s$^{-1}$ level was found, although line broadening at that
level was needed to explain the data \citep{Hetal99}. The previous
high resolution spectroscopy studies of main sequence stars in the
halo resulted in observational limits on their rotation rate of 8 km
s$^{-1}$ \citep{CP81}.  We also have stronger constraints on the
timescale for internal angular momentum transport in main sequence
stars from studies of young open clusters \citep[and references
therein]{SPT99}. Finally, we now have observational data for stars on
the horizontal branch hotter than 12 000 K in M13, which show a
markedly different behavior from the cooler horizontal branch stars in
that cluster -- they are all rotating at less than 10 km $s^{-1}$
\citep{behr99}. This difference in behavior of horizontal branch stars
of different masses could prove enlightening for questions of
horizontal branch morphology, mass loss on the giant branch, and
angular momentum evolution during the horizontal branch phase.

In this paper, we wish to explore the following question: During the
evolution of a star from the main sequence turnoff to the horizontal
branch, how can the star retain a large amount of angular momentum
under a relatively limited angular momentum budget and with
substantial mass (and hence angular momentum) loss?  We begin with a
star at the turnoff which is rotating as a solid body, and explore the
predicted range of horizontal branch rotation rates as a function of
$T_{eff}$ by considering limiting case assumptions about the rotation
law in the giant star convection zone (solid body or constant specific
angular momentum), internal angular momentum transport in giants
(angular momentum deposited in the core is retained or all angular
momentum is in the convective envelope), and internal angular momentum
transport on the horizontal branch (solid body rotation or local
conservation of angular momentum between the giant branch tip and the
horizontal branch).  In section 2, we outline the method we used to
construct our stellar models. We present the results in section 3, and
discuss the implications in section 4. A summary of our conclusions is
given in section 5.

\section{METHOD}

\subsection{The Evolutionary Models}

We used the Yale Rotating Evolution Code (YREC, see \cite{SPT99}) to
calculate a standard non-rotating stellar model from the main sequence
up the giant branch to the helium core flash. This model was chosen to
have parameters appropriate for stars in M13: M=0.8 M$_{\odot}$,
Z=0.0006 ([Fe/H]=-1.5) and Y=0.23. The turnoff age of this star is
14.7 Gyr. The mixing length parameter, set by calibrating a solar
model, was 1.7. We used OPAL opacities \citep{IR96} for temperatures
greater than $\log T = 4.0$, \cite{AF94} opacities for lower
temperatures, the Saha equation of state, and grey Eddington
atmospheres. The choice of the equation of state and atmospheres was
made necessary by the range of evolutionary stages that we are
investigating. The more recent, and more accurate, equations of state
\citep{RSI96,SCV95} and atmospheres \citep{AH95} unfortunately do not
yet extend to the temperatures and densities required for stellar
models near the tip of the giant branch. However, since this work is
an initial study of rotational evolution in these advanced phases, the
qualitative results presented here will not be affected by minor
changes in the position of the evolutionary track in the HR diagram.

\subsection{Initial Angular Momentum Budget}

Helioseismology can be used to infer the internal rotation profile of
the Sun by observing the rotational splitting of solar p-modes. The
results imply that the Sun's radiative core is rotating as a solid
body down to about $R=0.2 R_{\sun}$, with some disagreement about
deeper layers \citep{Chap99,Laz96,Cor98}.  The angular velocity of the
solar surface convection zone is dependent on latitude but not on
radius \citep{T96}.  Old metal-poor stars do not have similar direct
constraints on their internal rotation, but the solar case is
certainly a good first approximation.  We therefore assume solid body
rotation throughout the interior of the main sequence turnoff
progenitor and examine whether it is possible to retain sufficient
angular momentum to explain the rapid observed horizontal branch
rotation rates.

Our initial conditions therefore reduce to the total moment of inertia
at the main sequence turnoff and the surface rotation rate.  An
extrapolation of the angular momentum loss model for Pop I stars
\citep{SPT99} yields $v_{surf} = 1$ km s$^{-1}$ at the turnoff and a
total angular momentum of 5 $\times 10^{47}$ g cm$^2$ s$^{-1}$,
compared with 2 $\times 10^{48}$ g cm$^2$ s$^{-1}$ for a star rotating
at the current observational limit of 4 km s$^{-1}$ (and is comparable
to the angular momentum of a solar model which rotates as a solid
body). We will therefore consider an initial angular momentum of $5
\times 10^{47}$ g cm$^2$ s$^{-1}$ as our base case and $2 \times
10^{48}$ g cm$^2$ s$^{-1}$ as our limiting case, with angular momentum
distributed as a function of mass the same as in a solid body rotator.

\subsection{Mass and Angular Momentum Loss on the Giant Branch}

In main sequence stars there is efficient angular momentum loss from a
magnetic wind.  Because of the low predicted surface rotation rates,
the amount of angular momentum loss from a magnetic wind is expected
to be small on the giant branch.  However, large amounts of mass loss
on the first ascent giant branch are required to explain the
distribution of horizontal branch star masses \citep{R73,LDZ90}, and
the matter at the surface will carry away at least its own local
angular momentum.  Mass and angular momentum loss was therefore
calculated as follows.

Mass loss on the giant branch was calculated using Reimers' formulation \citep{R75}:
\begin{equation}
\dot{M} = \alpha \frac{L}{gR} M_{\sun} yr^{-1},
\end{equation}
where $L$ is the total luminosity, $g$ is the surface gravity, and $R$
is the radius of the star. The constant $\alpha$ can be varied to
allow for different amounts of mass to be lost over the entire giant
branch evolution.  The mass loss was assumed to begin at the point of
maximum convection zone depth in mass (see section 2.4). The loss of
mass was taken into account in the evolution of the star.

We calculated the angular momentum lost from the star as it loses
mass, following the method used in \cite{PDD}. Rotation is not
included in the structural evolution of the star, and rotational
mixing is not considered in these models. As mass is lost from the
star, the fractional angular momentum loss in a given timestep is
\begin{equation}
\Delta J = \frac{2}{3} \Delta M R^2 \omega
\end{equation}
where $\Delta M$ is the total mass lost during the time step, $R$ is
the stellar radius, and $\omega$ is the rotation rate at the surface of the
star. Models with different mass loss rates will therefore experience
different amounts of angular momentum loss.

The Reimers' mass loss constant $\alpha$ was varied so that different
amounts of mass, up to almost all of the mass outside the helium
burning core, were lost from the initially 0.8 M$_{\odot}$ model. For
each of the resulting combinations of core mass (taken from the model
at the tip of the giant branch) and envelope mass, we calculated zero
age horizontal branch evolutionary models and determined the stars'
temperature, radius, total moment of inertia, and moment of inertia of
the convection zone on the horizontal branch. The structural
properties of each horizontal branch model are given in table 1.

The total amount of angular momentum that survives in the convective
envelope and radiative core depends not only upon the amount of mass
and angular momentum loss, but also upon the efficiency of internal
angular momentum transport from the contracting radiative core into
the expanding convective envelope, and the angular momentum
distribution within the convective envelope. We outline below the
limiting cases we have chosen to investigate the different
possibilities.

\subsection{Internal Redistribution of Angular 
Momentum from the Turnoff to the Giant Branch Tip}

The radiative core on the giant branch is a potential reservoir of
angular momentum which could be redistributed to the surface layers
after the helium flash.  There is a significant fraction of the
initial mass (0.313 $M_{\sun}$) which is never incorporated in the
outer convective envelope.  This inner core contains 16.9\% of the
moment of inertia, and therefore angular momentum, of the main
sequence turnoff precursor assuming solid body rotation.  Another
0.156 $M_{\sun}$ from the convective envelope falls into the radiative
interior between the point of maximum depth in mass and the helium
flash; the angular momentum content of this component depends on the
angular momentum distribution in convective regions.

We consider two limiting cases for the angular momentum content of the
radiative core on the giant branch: 

\begin{itemize}

\item Angular momentum present in the core at the point of maximum
convection zone depth or deposited in the radiative core remains there
until the tip of the giant branch.

This assumption preserves the maximum amount of core angular momentum
unless angular momentum is pumped into the rapidly rotating core from
the slowly rotating envelope.  In this first case we assume that
angular momentum is conserved locally in the radiative interior.  This
is not necessarily in contradiction with the small degree of internal
solar differential rotation inferred from helioseismology given the
much shorter time scale for upper giant branch evolution (18 Myr from
the point of maximum convection zone depth to the tip of the RGB, for
example) compared with the age of the Sun (4.57 Gyr).  We also note
that studies of angular momentum evolution in young open cluster stars
also require a time scale for internal angular momentum transport
between 20 and 100 Myr \citep[and references therein]{KMC95, A98, SPT99}.

\item Solid body rotation at all times
in the radiative core with $\omega$ = $\omega_{CZbase}$.

The second assumption is the limit of instantaneous angular momentum
transport within the radiative core; this would correspond to solid
body rotation in the radiative core of the interior with an angular
velocity equal to that of the base of the surface convection zone.
Given the small fraction of the total moment of inertia in the
radiative core of a giant branch star (less than 1\% at the giant
branch tip) this is effectively equivalent to depositing all of the
angular momentum in the convective envelope.

\end{itemize}

\subsection {Rotation Law in Convective Regions}

The principal observational constraint on the angular momentum
distribution in convective regions is from helioseismology: the
angular velocity is independent of radius in the solar convection
zone. However, rotation is much slower on the giant branch, which
implies that this may not hold across the large number of pressure
scale heights in a giant branch convective envelope. We therefore
consider two limiting cases:

\begin{itemize}

\item Solid body rotation at all times in convective
regions

The assumption of solid body rotation at all times in convective
regions concentrates high specific angular momentum material in the
outer layers of the convection zone. When mass is lost from the star,
a higher fraction of the angular momentum of the star goes with it
than would be the case if the deeper layers had a larger fraction of
the angular momentum.  This assumption implies very slow rotation at
the base of the convection zone for reasonable initial rotation
rates. It is therefore very difficult to drive local mixing in the
radiative interior \citep{SM79}.

\item Constant specific angular momentum at all times in convective
regions

The assumption of constant specific angular momentum concentrates the
envelope angular momentum in the deeper layers of the envelope, with
the bulk of the mass.  This minimizes the fraction of the envelope
angular momentum carried away by mass loss.

\end{itemize}

\subsection{Internal Redistribution of Angular Momentum on the
Horizontal Branch}

As a star undergoes the helium flash, its core expands and its
envelope contracts, essentially reversing the structural effects of
giant branch evolution. Therefore, any steep angular velocity profile
which has been set up by the giant branch evolution will be
smoothed. It is also possible that hydrodynamic mechanisms, triggered
by the helium flash, will redistribute angular momentum throughout the
star on very short timescales. Therefore, we consider the following
two limiting cases:

\begin{itemize}
\item Solid body rotation of the entire star on the horizontal branch

This case corresponds to the maximum redistribution of the 
angular momentum from the core throughout the star. The rotation rate
at the surface of the horizontal branch star is given by $\omega =
J_{tot}/I_{tot}$.  This permits angular momentum preserved in the
radiative core of the giant branch precursor to be redistributed to
the outer layers of the horizontal branch star.

\item Local conservation of angular momentum between the giant branch
tip and the horizontal branch.

In this case, no angular momentum is redistributed during the helium
flash. The specific angular momentum profile of the giant branch star
is simply carried to the new structure of the star on the horizontal
branch. The surface rotation rate is given by the specific angular
momentum at the surface of the giant branch star and the radius of
the star on the zero age horizontal branch.  In the limit of a very
thin surface convective zone on the horizontal branch, the ratio of
the horizontal branch to the giant branch tip rotation velocities is
equal to the radius at the giant branch tip divided by the horizontal
branch star radius.

\end{itemize}

\subsection{Detailed Description of Individual Cases}

The ultimate goal of these calculations is the rotational velocity on
the horizontal branch. First, we need to make an assumption about the
total amount of angular momentum available at the turnoff, as
discussed above. Second, we calculate the rotation rate on the
horizontal branch according to one of two scenarios. If we assume
local conservation of angular momentum between the tip of the giant
branch and the horizontal branch, then the amount of specific angular
momentum on the horizontal branch is equal to the specific angular
momentum at the tip of the giant branch, and
\begin{equation}
J_M(HB)=\frac{2}{3} \omega_{HB} R_{HB}^2 = J_M(tip)
\end{equation}
so that
\begin{equation}
v_{HB}=\frac{3 J_M(tip)}{2 R_{HB}}.
\end{equation}
The other extreme is to assume that the horizontal branch star rotates
as a solid body. In this case, the total angular momentum on the
horizontal branch is equal to that at the giant branch tip, and 
\begin{equation}
v_{HB}=\frac{J_{tot}(tip)}{I_{tot}(HB)} R_{HB}.
\end{equation}
Therefore, we need four pieces of information from our models: the
total angular momentum of the star at the giant branch tip, the
surface specific angular momentum at the giant branch tip, the moment
of inertia of the star on the horizontal branch, and the radius of the
star on the horizontal branch. The moment of inertia and the radius
are given by the horizontal branch model. The total angular momentum
and the surface value of the angular momentum per unit mass are
derived as outlined below for each of the four giant branch angular
momentum distributions.

Case A is the case in which the entire star always rotates as a solid
body. The angular velocity $\omega$ is known at all times from
$J_{tot} = \omega I_{tot}$, where $I_{tot}$ is the total moment of
inertia of the star. Therefore, following \cite{PDD}, the fractional
angular momentum loss in a given timestep is
\begin{equation}
\frac{\Delta J}{J_{tot}} = \frac{2}{3} \frac{\Delta M R^2}{I_{tot}},
\end{equation}
where $\Delta M$ is the total mass lost during the time step, $R$ is
the radius, and $I_{tot}$ is the total moment of inertia of the star.
The total angular momentum lost over the entire giant branch evolution
is the sum of the individual $\Delta J$, and the specific angular
momentum at the surface is given by $J_M=2/3 \omega R^2$

In case B, we assume that the convection zone rotates as a solid body,
and that the core retains the angular momentum that it had on the main
sequence. The initial angular momentum in the core is given by
$J_{core}= \omega(MS) I_{core}(MS)$, where $I_{core}$ is the moment of
inertia of the portion of the star on the main sequence that is
contained in the radiative interior where the surface convection zone
reaches its maximum depth. The amount of angular momentum in the
envelope at any given time is determined by two things: the amount of
mass lost from the surface of the star ($\Delta M_1$), and the amount
of mass which becomes part of the core as the convection zone recedes
($\Delta M_2$). The total angular momentum of the envelope at any
given timestep is therefore
\begin{equation}
J_{env} = J_{env,old} - \frac{2}{3} \omega_{env} R_{surface}^2
\Delta M_1 - \frac{2}{3} \omega_{env} R_{CZ base}^2 \Delta M_2
\end{equation}
The amount of angular momentum which moves from the envelope to the
core as the convection zone recedes is much less than 1\% of the
total, and so we neglected the term second term in equation
7. Therefore, the amount of angular momentum lost during a given
timestep is given by equation 6, but with $I_{tot}$ replaced with
$I_{env}$ and $J_{tot}$ replaced with $J_{env}$. The total angular
momentum of this star at the tip of the giant branch is equal to the
amount of angular momentum maintained in the envelope, plus the
initial angular momentum of the core, and the specific angular
momentum at the giant branch tip is given by $J_M=2/3 \omega
R_{tip}^2$.

Case C is the case in which the core of the star rotates as a solid
body at all times with $\omega = \omega_{CZ base}$ , and the
convection zone has constant specific angular momentum. Therefore, the
amount of angular momentum per unit mass changes as the star loses
mass, and as the convection zone depth changes due to structural
changes in the giant. The total angular momentum in the core is given
by
\begin{equation}
J_{core}=\frac{3 I_{core} J_M(env)}{2 R_{CZ base}^2}
\end{equation}
where $J_M(env)$ is the specific angular momentum of the convection
zone.  Since the moment of inertia of the core is much smaller than
the total moment of inertia of the star as a whole (less than $2
\times 10^{-5} J_{tot}$) we assumed that the core retained no angular
momentum at all, and therefore the specific angular momentum in the
convection zone is given by $J_M = J_{tot}/M_{env}$, where $M_{env}$
is the current mass of the convection zone. As the star loses mass,
the amount of total angular momentum at each timestep is simply
$J_{tot}=J_{tot,old} - J_M(env) \Delta M$. The total angular momentum of
the star at the tip of the giant branch is the sum of the losses at
each timestep.

The final case, case D, is the simplest in which to calculate the
amount of angular momentum left at the tip of the giant branch. In
this case, the core of the star (as defined in case B) retains its
main sequence angular momentum, and the envelope has constant specific
angular momentum. The value of $J_M$ is set at the beginning of the
calculation to be equation  to $J_{env}/M_{env}$, where both
quantities are evaluated at the point of maximum convection zone
depth, and is not allowed to change throughout the subsequent
evolution. In this case, the total amount of angular momentum lost
over the entire giant branch evolution is $\Delta J_{tot} = J_M
\Delta M_{tot}$. 

\section{RESULTS}

\subsection{Angular Momentum Evolution from the Turnoff to the
Horizontal Branch}

Figure 1 shows the angular momentum evolution of a star from the main
sequence to the horizontal branch, demonstrating the effects of the
changing structure on the rotation rate of the star. In each panel, we
show the angular velocity profile as a function of radius in the
star. We have also marked the positions of the center of the hydrogen
burning shell, the radius at which the hydrogen content is 50\% by
mass, and the base of the surface convection zone. The star shown here
has a mass of 0.8 $M_{\sun}$ and we assume no mass loss or angular
momentum loss along the giant branch, and no internal transport of
angular momentum. All angular momentum
evolution is caused by the changing structure of the star.  The first
panel shows a star at the main sequence turnoff, which we have assumed
is rotating as a solid body ($\omega$ = constant) with radius. The
second panel shows the rotation profile of the star at the position on
the giant branch where the convection zone reaches its maximum depth
in mass. The profile at the tip of the giant branch is shown in the
third panel, and the final panel gives the rotational profile on the
zero age horizontal branch.  The surface rotational velocities and
ages of the star are given in each panel.

As the star evolves along the subgiant branch and up the giant branch
to the position of the maximum convection zone depth, the core
contracts, the convective envelope deepens, and the surface
expands. The angular momentum which is found in the convection zone is
redistributed according to the rotation law for convection regions (in
this case assumed to be solid body), and the core of the star spins up
as it contracts. This trend continues for the rest of the giant branch
evolution, up to the tip of the giant branch. As the core contracts,
it continues to spin up, while material at the base of the convection
zone falls into the hydrogen burning shell, and the surface continues
to expand. At the tip of the giant branch, the surface of the star is
rotating 100 times slower than it was on the main sequence.  During
the helium flash, when the star moves from the tip of the giant branch
to its position on the horizontal branch, the giant branch evolution
essentially runs in reverse. The surface of the star contracts, and
the core of the star expands, flattening the rotational profile and
raising the surface rotation rate by a factor of $\sim$ 10. The
horizontal branch star is not rotating as a solid body, but it is much
closer to solid body rotation than the highly differentially rotating
giant branch stars.

Figure 2 shows the impact of mass loss on the evolution of the moment
of inertia of the star along the giant branch. We have plotted total
moment of inertia of the star as a function of $\log (L/L_{\sun})$ for
evolutionary tracks with different amounts of total mass loss. Low on
the giant branch, the star does not lose much mass in each timestep,
and so the stars are not significantly different from each other. Near
the tip of the giant branch, however, the amount of mass lost from the
star increases, which reduces the size of the giant convection zone,
resulting in a smaller star and hence a smaller moment of inertia than
predicted from non-mass-losing models. The evolutionary tracks used in
this paper represent an improvement over the work done in \cite{PDD},
in which mass loss was not included in the evolutionary tracks, and
all results were based on a track like the 0.8 $M_{\sun}$ track shown
in figure 2. Under the assumption that the moment of inertia was not
significantly affected by the mass loss, \cite{PDD} predicted too little
angular momentum loss, particularly for the hotter horizontal branch
stars.

\subsection{Parameter Variations at Fixed Horizontal Branch Mass}

We now examine the impact of different assumptions about internal
angular momentum transport after the main sequence turnoff, the
rotation profile enforced in convective regions, and internal angular
momentum transport during the helium flash and on the horizontal
branch.

In Figures 3 and 4 we illustrate the specific angular momentum as a
function of mass at different epochs for two different choices of the
rotation law in convective regions.  In each figure the top panel
corresponds to the case of local angular momentum conservation in the
radiative core while the bottom set of panels corresponds to the case
of solid body rotation at the angular velocity of the base of the
surface convection zone.  We have picked a reference horizontal branch
mass of 0.6 $M_{\sun}$, corresponding to an effective temperature on
the horizontal branch of $\sim$ 10 000 K and total mass loss on the
giant branch of 0.2 $M_{\sun}$ .  We now discuss results for solid
body rotation in convective regions (Figure 3) and uniform specific
angular momentum in convective regions (Figure 4) in turn.

The assumption of uniform rotation in convective regions implies that
the matter at the surface will have relatively high specific angular
momentum; as a result, mass loss will effectively drain angular
momentum from the envelope (Figure 3).  Furthermore, the base of the
surface convection zone will have very low specific angular momentum;
this minimizes the angular momentum content of the radiative core even
if the core angular momentum is not redistributed to the envelope.

Our reference model at 10 000 K retains from 2\% to 18.4\% of its
turnoff angular momentum, depending on whether solid-body rotation is
enforced in the core or whether the angular momentum deposited in the
core remains there.  Once on the horizontal branch, the surface
rotation rate will depend upon whether angular momentum from the core
is redistributed to the envelope.  For local conservation of angular
momentum from the giant branch tip to the horizontal branch, the
predicted surface rotation rates are extremely low, of order 0.01 km
s$^{-1}$.  If angular momentum is redistributed from the core to the
envelope on the horizontal branch the surface rotation rate for our
reference model is in the range 0.24 km s$^{-1}$ to 2.23 km s$^{-1}$
depending upon whether or not angular momentum was preserved in the
core on the giant branch. The expected surface rotation
rates on the horizontal branch for models in this class will be low
for all assumptions about internal angular momentum transport on the
giant and horizontal branch. 

If there is constant specific angular momentum in giant branch
convective envelopes, there are two effects that limit the amount of
envelope angular momentum loss.  First, the fraction of the envelope
angular momentum which is lost is proportional to either the fraction
of the mass above the point of maximum convection zone depth (local
conservation of angular momentum in the core) or to the fraction of
the convective envelope which is lost (solid body rotation in the
radiative core).  There will therefore be more angular momentum
available on the horizontal branch for this case (Figure 4). Our
reference model retains from 49.8\% to 69.4\% of its turnoff angular
momentum for models with solid body core rotation and local angular
momentum conservation in the core respectively.  The inferred surface
rotation rates on the horizontal branch range from a low of 0.01 km
s$^{-1}$ for local conservation of angular momentum from the giant
branch tip to the horizontal branch to a range of 6.04 - 8.42 km
s$^{-1}$ if there is horizontal branch angular momentum
redistribution.

All of these cases are well below the observed rotation rates of blue
horizontal branch stars, indicating that for solid body main sequence
rotation a higher main sequence rotation rate than 1 km s$^{-1}$ is
needed to explain the data.  In the case of solid-body rotation in the
convective envelope on the giant branch, the required main sequence
rotation rates are prohibitively high; a similar conclusion was
reached in \cite{PDD}.

\subsection{Trends with Mass on the Horizontal Branch}

Different amounts of giant branch mass loss will produce horizontal
branch models with different $T_{eff}$; models which lose more mass
will also lose more angular momentum.  Bluer horizontal branch models
will therefore have systematically less total angular momentum than
redder horizontal branch models.  However, the moment of inertia
decreases strongly for the bluer horizontal branch models, and it is
therefore not clear a priori that lower surface rotation velocities
would be predicted.

The overall structural properties of our horizontal branch models are
summarized in Table 1.  All models are shown at the zero age
horizontal branch.  There is a strong trend to decreased moment of
inertia for the bluer models, which can be easily understood
physically.  The bulk of the mass is in the core, while the bulk of
the moment of inertia is in the envelope.  The small change in the
total mass at the blue end of the horizontal branch corresponds to a
large fractional change in the mass and moment of inertia of the
envelope, but does not affect the mass of the core.

The fraction of the angular momentum retained under our different
cases is illustrated as a function of mass lost on the giant branch
branch in Figure 5; we also present this data in Table 2.  The
predictions of the different classes of models for the amount of
available angular momentum diverge.  In the pure solid body model
(case A), virtually all of the angular momentum is lost for high giant
branch mass loss rates.  Local conservation of angular momentum in the
core and solid body rotation in the envelope (case B) results in a
minimum level of angular momentum (retained in the core) even when
essentially all of the envelope angular momentum is lost.  The two
cases with constant specific angular momentum in the convective
envelope systematically retain a higher fraction of their turnoff
angular momentum; again, the case where a reservoir of angular
momentum is retained in the giant branch core (case D) leaves more
angular momentum than a set of models with solid-body core rotation on
the giant branch (case C).

Horizontal branch surface rotation velocities can be inferred from the
angular momentum content and the moment of inertia and radius as a
function of $T_{eff}$ (Table 3.)  All of the cases with local
conservation of angular momentum from the giant branch tip to the
horizontal branch have extremely low surface rotation rates (less than
1 km s$^{-1}$).  There is a range of higher rotation rates if there is
effective angular momentum redistribution during the horizontal branch
phase of evolution.  We therefore conclude that a range of surface
rotation rates from under 1 km s$^{-1}$ to a ceiling dependent on the
prior evolution can be generated on the horizontal branch, and the
observed range in rotation rates could reflect evolution on the
horizontal branch itself rather than an intrinsic range of initial
rotation rates at fixed giant branch tip mass.  We will therefore
treat the surface rotation rates for solid body horizontal branch
rotation as an upper envelope; a range of rotation rates up to this
level are consistent with the physics of the different cases that we
have considered.

We compare the inferred surface rotation rates as a function of
$T_{eff}$ with the observational data in Figures 6 and 7.  Figure 6 is
for a turnoff rotation velocity of 1 km s$^{-1}$, consistent with an
extrapolation of the Population I angular momentum loss law to
Population II stars; Figure 7 is the same set of cases for a turnoff
rotation velocity of 4 km s$^{-1}$, at the level where additional line
broadening is observed in high-precision spectroscopic studies of main
sequence halo stars. We note that the upper envelope of the
observations in the 4 km s$^{-1}$ case is well-matched in the cool
horizontal branch stars (T$_{eff} <$ 10 000 K), but we predict no
decrease in rotational velocities for hotter stars, and no bimodal or
sharp cutoff behavior.

\section{DISCUSSION}

Main sequence metal-poor stars rotate slowly, while some metal-poor
horizontal branch stars are rapid rotators.  This combination requires
one of the following:  

\begin{enumerate}
\item  Main sequence stars have a larger angular momentum content than
inferred from their slow surface rotation, i.e. they possess rapidly
rotating cores; or
\item A significant fraction of a limited angular momentum budget
survives extensive giant branch mass loss; or
\item There is a source of angular momentum for the rapid rotators on the
giant branch or during the helium flash.
\end{enumerate}

We believe that the second explanation is the most likely one, and
that by extension the angular momentum distribution and evolution in
evolved stars is very different than that inferred for the Sun.  Our
results also provide some insight into the role of internal angular
momentum transport on the horizontal branch and both the origin of the
(real) spread in rotation rates at fixed effective temperature and the
surprising temperature dependence of horizontal branch rotation that has been
observed recently.  We address these issues individually below.

\subsection{Internal Main Sequence Rotation}

The underlying problem that we are addressing is not new. \cite{PDD}
compared the horizontal branch rotation measurements of
\cite{P83,P85a} with different theoretical models and concluded that
there were two classes of possible solutions: rapid main sequence core
rotation or differential rotation with depth in giant branch
convective envelopes.  The strongest other constraints on the internal
angular momentum distribution in stars are currently obtained from
helioseismology. In 1991, helioseismic inversions were precise
enough to rule out strong differential rotation with depth in the
solar convection zone but not in the radiative core; therefore, a
rapidly rotating core on the main sequence was the favored solution.
The advent of more precise helioseismic determinations of the internal
solar rotation, however, leads to a fundamental contradiction: either
the core rotation of metal-poor main sequence stars is fundamentally
different from the solar case or the convective envelopes of evolved
stars are different from the solar convective envelope.  A measurement
of the surface rotation rates of main sequence turnoff metal-poor
stars will aid in constraining the total amount of angular momentum
available during giant branch evolution.

Our principle result is that a solar-like internal rotation profile on
the main sequence can only be reconciled with rapid horizontal branch
rotation and slow main sequence rotation if the giant branch evolution
in both the convective envelope and the radiative core is very
different during these three evolutionary phases. The one class of
models which could reproduce the observed upper envelope of horizontal
branch rotation rates while remaining consistent with the observed
limits on main sequence rotation was the case with constant specific
angular momentum in giant branch convective envelopes, retention of a
rapidly rotating core on the giant branch, and subsequent angular
momentum redistribution from the core to the envelope on the
horizontal branch.  Furthermore, even this case of models requires
significantly more rapid surface main sequence rotation ($\sim$ 4 km
s$^{-1}$) than predicted by an extrapolation of the Population I
angular momentum loss law to Population II stars ($\sim$ 1
km s$^{-1}$).

There is a possible explanation for metal-poor turnoff stars having
higher rotation rates than expected from a straightforward
extrapolation of the Population I angular momentum loss law.  This
would be related to the Population II analog of the observed
Population I break in the rotational properties of high and low mass
stars \citep{K65}: low mass stars experience angular momentum loss and
high mass stars do not. \cite{DL78} used a Rossby scaling argument as follows.
There exists a critical rotation rate for a given convective overturn
time scale below which a dynamo does not operate.  As the convective
overturn time scale decreases the critical rotation rate rises; there
will therefore be a transition region from low mass stars that
experience angular momentum loss for their entire main sequence
lifetime to higher mass stars that only experience spindown until
their surface rotation rate drops below the critical threshold.  The
convective envelopes of metal-poor turnoff stars are also thin, which
suggests that they might experience less angular momentum loss than
would be obtained from a prescription without a critical threshold.
The convective overturn time scale is a strong function of $T_{eff}$ in
the F star regime and also depends on metal abundance.  This opens up
the further possibility that the available angular momentum budget
could depend strongly on metal abundance and age, which could lead to
large variations in the degree of rotation in evolved stars and the
degree of rotational mixing on the giant branch from modest abundance
and age variations.

\subsection{Angular Momentum Evolution on the Giant Branch}

If a solar-like internal angular momentum distribution (rotation
weakly dependent on depth) is postulated for main sequence metal-poor
stars, strong differential rotation in both the envelopes and cores of
giants is required to produce rapid horizontal branch rotation.  This
is not necessarily a contradiction given the very different structure
and time scale for evolution.

The internal rotation of the Sun as deduced from helioseismology is
straightforward.  The solar convection zone has differential rotation
with latitude but not with depth; there is a shear layer below the
convection zone where latitudinal differential rotation vanishes, and
below the shear layer the internal solar rotation appears to be nearly
independent of both depth and latitude.  Because of the short
convective overturn time scale in the Sun, a good theoretical case can
be made that the properties of the solar convection zone should apply
generally when the convective and rotational velocities are of the
same order.  However, it is far from clear that the same conclusion
should be obtained for all convective regions in all stars regardless
of the size and rotation rate.

Both the surface and convection zone base rotation velocities are
extremely small in giant branch envelopes because of the small angular
momentum reservoir and large moment of inertia.  At the point of
maximum convection zone depth the rotation velocities at the
convection zone base and surface are respectively 0.01 and 0.07 km
s$^{-1}$ for an initial angular momentum reservoir of $5 \times
10^{47}$ g cm$^2$ s$^{-1}$, solid body rotation in the convective
envelope and no angular momentum loss.  At the giant branch tip these
rotation velocities are, respectively, 0.0002 and 0.009 km s$^{-1}$.
These velocities are small in comparison with typical convective
velocities.  By comparison, the solar rotation velocity of 2 km
s$^{-1}$ is of the same order or larger than the convective velocity.
Therefore, in the Sun, differential rotation with depth would induce a
large absolute shear relative to the typical turbulent velocity.  In a
giant branch star, however, the convective velocities are so much
larger than the rotational velocities that even a large relative
gradient in $\omega$ is a small absolute gradient in $\omega$. It is
therefore possible that a giant branch convection zone, even with
constant specific angular momentum, would have very little shear, and
need not rotate as a solid body.

The case of the radiative core is even more complex: the solar data by
itself requires effective angular momentum transport by the age of the
Sun and by extension also in main sequence globular cluster stars.
However, the solar data does not necessarily require efficient angular
momentum transport in radiative regions on much shorter time scales.
Several different studies of the angular momentum evolution of low
mass stars have concluded that the time scale for the coupling of the
radiative core and the convective envelope is in the range of 20-100
Myr \citep{KMC95,A98,SPT99}.  The evolutionary time scale on the giant
branch is comparable and the degree of differential rotation generated
by structural change on the giant branch is much greater than the
differential rotation generated by main sequence angular momentum
loss.

It should be noted that the conditions needed to explain rapid
horizontal branch rotation (i.e. differential rotation on the giant
branch) are the same as those required to produce mixing on the giant
branch \citep{SM79,P97}. This is an encouraging sign for the
possibility of self-consistent giant branch mixing models.

\subsection{Angular Momentum Evolution on the Horizontal Branch}

In all of the classes of models that we have examined, horizontal
branch rotation rates of 1 km s$^{-1}$ or less are predicted if the
giant branch angular momentum distribution is evolved to the
horizontal branch using local conservation of angular momentum.  To
first order, horizontal branch rotation rates of order the main
sequence turnoff rotation rate or less are predicted.  If angular
momentum from the core can be redistributed to the envelope more rapid
rotation on the horizontal branch - up to 40 km s$^{-1}$ - is possible
for a main sequence turnoff rotation rate of 4 km s$^{-1}$.  The
detection of rotation at the 40 km s$^{-1}$ level on the horizontal
branch therefore requires internal angular momentum transport either
during the helium flash or on the horizontal branch itself.

Hydrodynamic simulations of the helium flash generally predict that
the core can become mixed but the whole star is not \citep{D96}; in
fact, the survival of a helium core is required to generate a
horizontal branch star rather than a helium-rich main sequence star.
The short time scale for evolution between the giant branch tip and
the horizontal branch also argues against a large-scale readjustment
of the angular momentum profile prior to the actual horizontal branch
evolution.  It therefore appears likely that ongoing angular momentum
transport during the horizontal branch phase will produce systematic
changes in the surface rotation rate as a function of time; the sense
of these changes would be that newly arrived stars on the horizontal
branch would have the lowest rotation rate and the more evolved stars
would have higher rotation rates.  We also note that in the absence of
effective angular momentum transport from core to envelope the surface
rotation rates of horizontal branch stars are predicted to be very
low.  These properties have interesting consequences for the
interpretation of both the observed range of rotation rates at fixed
effective temperature and the difference in the observed rotation
rates of hotter and cooler horizontal branch stars.

\subsubsection{The Range of Horizontal Branch Rotation Rates at Fixed 
$T_{eff}$}

The horizontal branch rotation data requires a range in rotation
velocity at fixed $T_{eff}$, which is contrary to expectations from an
extrapolation of trends observed on the main sequence.  We believe
that angular momentum evolution on the horizontal branch is
responsible for this range, rather than a true range of total angular
momenta at fixed horizontal branch mass.  The range of observed
rotation rates at fixed mass and composition is observed to decrease
strongly with increased age; e.g., compare the distribution of
rotation rates in the Hyades cluster of \cite{R87} with the
distribution of rotation rates in the Pleiades cluster \citep{S93}.
This is a natural consequence of an angular momentum loss rate that
increases with $\omega$.  Different theoretical models can predict
different internal angular momentum distributions in turnoff stars,
but the memory of the initial conditions would be expected to be
erased if either hydrodynamic mechanisms or internal gravity waves are
principally responsible for internal angular momentum transport in
stars.  The case of internal angular momentum transport by magnetic
fields \citep{KMC95} is different: the overall characteristics
depend on the overall internal magnetic field morphology, i.e. whether
magnetic field lines connect the convective envelope with the entire
radiative core.  This is observationally testable: the intrinsic
dispersion in rotation rates in clusters older than the Hyades could
be measured directly, and a dispersion in rotation rates at older ages
would be evidence of star-to-star differences in internal magnetic
field morphology.

Our limiting cases, however, indicate that there is another phenomenon
which must be accounted for when interpreting horizontal branch
rotation rates. The initial horizontal branch rotation will be low for
all cases. As a star evolves, internal angular momentum transport
could cause the star to rotate more like a solid body, on timescales
which could be comparable to the horizontal branch lifetime of these
stars. If this scenario were correct, then rotation rates of stars
should increase with luminosity at approximately fixed effective
temperature (i.e. along evolutionary tracks). It will be necessary to
identify those stars which are evolving in T$_{eff}$, on their way to
the asymptotic giant branch, so as not to confuse them with those
stars which have just started evolving off the zero age horizontal
branch.

\subsubsection{Trends in Rotation with $T_{eff}$ on the Horizontal Branch}

The models for horizontal branch stars cooler than $\sim$12 000 K
demonstrate a morphology consistent with the upper envelope of the
observed rotation rates of these stars (see figure 7), as long as the
initial angular momentum budget is high. Since the lowest temperature
stars retain most of their mass, and hence their moments of inertia,
they are rotating quite slowly. As temperature increases along the
horizontal branch, the stars lose more mass and therefore more angular
momentum, but their moments of inertia decrease more rapidly than the
angular momentum loss, and so the stars rotate faster as their
temperature increases.

However, the abrupt drop in rotation velocity at $\sim$12 000 K and
the apparent bimodal distribution of rotation rates along the
horizontal branch is not predicted by our models. Since we assume that
horizontal branch temperature is solely a function of horizontal branch
mass, we predict a smooth behavior of horizontal branch properties
with T$_{eff}$. Any explanation which invokes different mass loss
rates on the giant branch (caused by some second parameter perhaps)
must include an explanation of why a small change in mass results
in a significant change in rotational properties on the horizontal
branch.

There is an important additional piece of information about these
hotter horizontal branch stars: the stars hotter than 12000 K all show
evidence for gravitational settling \citep{behr99a}.  Gravitational
settling of heavy elements creates mean molecular weight gradients in
the outer parts of the star, which inhibits angular momentum transport
in the star (Pinsonneault 1997 and references therein, Vauclair
1999). This could be an indication that when the mean molecular weight
gradients become large enough, they prevent angular momentum
redistribution from the rapidly rotating core to the slowly rotating
envelope. Theory predicts that this should be a threshold process,
where the mean molecular weight gradient needs to reach a critical
value before the inhibition of angular momentum transport begins
\citep{M53,R91}. If this explanation is correct, the rotation boundary
in the horizontal branch should coincide with the gravitational
settling boundary in $T_{eff}$.

\subsection{Other Possibilities}

We have restricted ourselves to models with solid body main sequence
rotation.  As a result, the limited angular momentum budget combined with
relatively rapid horizontal branch rotation sets stringent limits on the
post-main sequence angular momentum evolution.  There are some other
interesting possibilities that should be mentioned.

First, there is the possibility that main sequence stars could contain
rapidly rotating cores.  Even the most recent helioseismic inversions
do not rule out rapid rotation in the deep solar interior, although
they also provide no support for the existence of rotation rapid
enough to contribute a large amount of angular momentum.  If some main
sequence stars retain rapidly rotating cores, the most likely
possibility is therefore that a range of internal rotation rates
survives in stars to late ages; in this case the internal rotation of
the Sun would simply be one of a range of possibilities.  This is
possible if stars have a range of overall magnetic field morphologies,
with the Sun being closer to a field that strongly couples the
radiative core and convective envelope and the progenitors of the
rapidly rotating horizontal branch stars presumably being stars with
weak coupling.  We note, however, that other mechanisms (hydrodynamic,
gravity waves) would still operate and that it is not clear that
different internal rotation can survive for a Hubble time even in this
case \citep{BCM99}.  This possibility would be made more likely if it
can be shown that the surface rotation of the main sequence precursors
to the horizontal branch stars have rotation rates lower than 4 km
s$^{-1}$. There is a second observational test as well: if a
dispersion in rotation rates at fixed mass, composition, and age is
found in turnoff stars in older systems this implies that the surface
rotation rates have not converged to the high degree predicted if
angular momentum transport is dominated by gravity waves or
hydrodynamic mechanisms.

Second, it is possible in a restricted set of circumstances for
angular momentum to be pumped from a slowly rotating envelope into a
more rapidly rotating core.  The equations for meridional circulation
as presented by \cite{Z91} involve the distortion from spherical
symmetry both locally (a centrifugal term) and globally (a quadrupole
term); furthermore, the transport of angular momentum involves both a
diffusive term and an advective term.  It is therefore possible, in
the limit of very slow local rotation and a highly distorted core, to
reverse the sign of the velocity and have a net transport of angular
momentum from the envelope into the core even if the average rotation
decreases with increased radius (Pinsonneault \& Sills, in
preparation).  This inversion requires a combination of rapid core and
slow envelope rotation, and it is not clear that a main sequence solid
body rotator will develop sufficient differential rotation with depth
to achieve this condition even for solid body rotation in the giant
branch convective envelope.  This is, however, an interesting
possibility that deserves future study.  In this case angular momentum
could be extracted from the envelope and deposited in the core; if the
quadrupole term is sufficiently large, it could even drive large-scale
mixing in the envelope of the giant branch star in the presence of
slow local rotation.

Thirdly, we cannot rule out the possibility of a source of angular
momentum on the giant branch itself.  A non-axisymmetric helium flash
is one possibility; another would be mass transfer between close
binaries, or between single stars and nearby giant planets
\citep{S98}.  The first case could potentially be tested by comparing
the velocity dispersion of horizontal branch stars relative to giants,
since we would expect that a non-axisymmetric helium flash should
impart linear as well as angular momentum to the horizontal branch
star. For the second case, mass and angular momentum transfer from a
companion is certainly possible; a merger of two stars either through
collisions or binary mergers is thought to be the origin of the blue
straggler phenomenon \citep{B95}.  However, one important problem with
these explanations is that the frequency of nearby brown dwarfs and
giant planets appears to be small, of order 5\% \citep{MBVS98}, while
the binary fraction in globular clusters is about 20\%
\citep{Hetal92}. Cluster-to-cluster differences in horizontal branch
rotation rates and the observed trends with $T_{eff}$ on the
horizontal branch would also have to be accounted for.

A possible explanation of the bimodal distribution in rotation rates
with $T_{eff}$ is to invoke two separate populations of stars on the
horizontal branch. The correlation between the sharp cutoff in
rotation rates with the existence of a gap in the horizontal branch of
M13 has prompted speculation that two populations are present in this
cluster, but the reason for these two populations is not yet
apparent. Dynamically created populations, such as evolved blue
stragglers or giant branch stars stripped by close encounters with
other stars, are candidates for this class of explanation. It is
difficult to understand, however, how the small populations of blue
stragglers or stars with binary systems could evolve to become as
numerous on the horizontal branch as the normal horizontal branch
stars in the cluster.

The work presented here is an initial study of some physically
interesting limiting cases. The simplified prescription of giant
branch evolution presented in this paper neglects some physical
effects which can change the structure and angular momentum profile of
the horizontal branch stars. Rotation also affects the structure of
the star, both directly and indirectly. Since rotation can provide an
additional source of support for the star, the core temperature of the
star will be lower than that of a non-rotating star. Therefore, the
star can live on the giant for longer until the core temperature
becomes high enough for the helium flash to occur. This will result in
a larger core mass than a non-rotating star, and will also increase
the amount of mass loss which can occur on the giant branch \citep{MG76}. The
internal transport of angular momentum during giant branch evolution
will be more complicated than the simple prescriptions presented in
this paper, and could also modify giant branch evolution enough to
change the position of the star in the HR diagram, the core mass at
helium flash and the amount of mass loss. We are planning on following
giant branch evolution, include the effects of rotation, using
machinery already present in the code, as well as considering angular
momentum transport by gravity waves, and advection.

\section{SUMMARY}

Stellar rotation is an important phenomenon, and there is a large and
growing body of data that is most easily interpreted as evidence for
rotationally induced mixing in stars..  The principal difficulty in
including rotation in theoretical stellar evolution calculations has
been the uncertainty about the internal angular momentum distribution
of stars.  We believe that the horizontal branch rotation data shed
some new and interesting light on this problem.  Evolved stars provide
complementary information to the extensive studies of rotation in the
Sun and low mass main sequence stars.

We have calculated simple models of angular momentum evolution on the
giant branch, concentrating on several limiting cases for the internal
angular momentum profiles on the giant branch and horizontal branch.
We find that rapid rotation on the horizontal branch can be reconciled
with solid body main sequence rotation if giant branch stars have
differential rotation in their convective envelopes and a rapidly
rotating core, which is then followed by a redistribution of angular
momentum on the horizontal branch. This set of conditions are the same
as the conditions which most favor mixing on the giant branch.  A
rotation rate of $\sim$ 4 km s$^{-1}$ at the main sequence turnoff is
required to explain the high rotation rates for the cooler horizontal
branch stars.

We suggest that the observed range in rotation rates for the cool
horizontal branch stars is a consequence of angular momentum evolution
on the horizontal branch. Angular momentum could be transported from
the core to the envelope over the horizontal branch lifetime of the
star, producing a correlation between rotation rate and luminosity on
the horizontal branch.  Our models do not predict the sharp cutoff in
rotation rates seen in the hot horizontal branch stars in M 13. We
suggest that this discontinuity is not tied to the angular momentum
loss rate from the giant branch, but rather could be a result of
gravitational settling. This creates a mean molecular gradient in the
star, which then inhibits angular momentum transport in the star.

We have proposed two observational tests pertaining to main sequence
stars which would be very useful in providing clarity to the problem
of angular momentum evolution on the giant branch. We need to know the
surface rotation rates of Population II main sequence stars, to
constrain the total angular momentum budget which is available for
loss and redistribution on the giant and horizontal branches. Also, by
studying the dispersion, or lack thereof, in rotational velocities of
main sequence stars in old open clusters, we can determine how the
internal magnetic field morphology could affect the dispersion in
total angular momenta in main sequence stars.

It would clarify the problem considerably to gather more observational
data on rotation rates of horizontal branch stars in globular
clusters. First, is the sharp cutoff in rotation rates seen in M 13
statistically significant? The data presented so far are convincing,
but only 7 of the approximately 25 stars hotter than 13 000 K have
been observed. Better statistics would eliminate this uncertainty.  It
would also be of great interest to observe the same kinds of stars in
other globular clusters. We already know that horizontal branch
morphology varies from one cluster to another. If we can determine
that this cutoff in rotation rates is ubiquitous, or alternatively is
correlated with horizontal branch morphology, that will provide a
valuable clue in determining the cause of the second parameter effect
and in understanding angular momentum evolution on the giant
branch. Secondly, we suggest that observers search for correlations
between rotation rates and surface abundances in horizontal branch
stars, to determine if there is a clear relationship between
gravitational settling and rotational properties on the horizontal
branch.

\acknowledgements This work was supported by NASA grant
NAG5-7150. A. S. wishes to recognize support from the Natural Sciences
and Engineering Research Council of Canada.

\clearpage

\begin{figure}
\figurenum{1}
\plotone{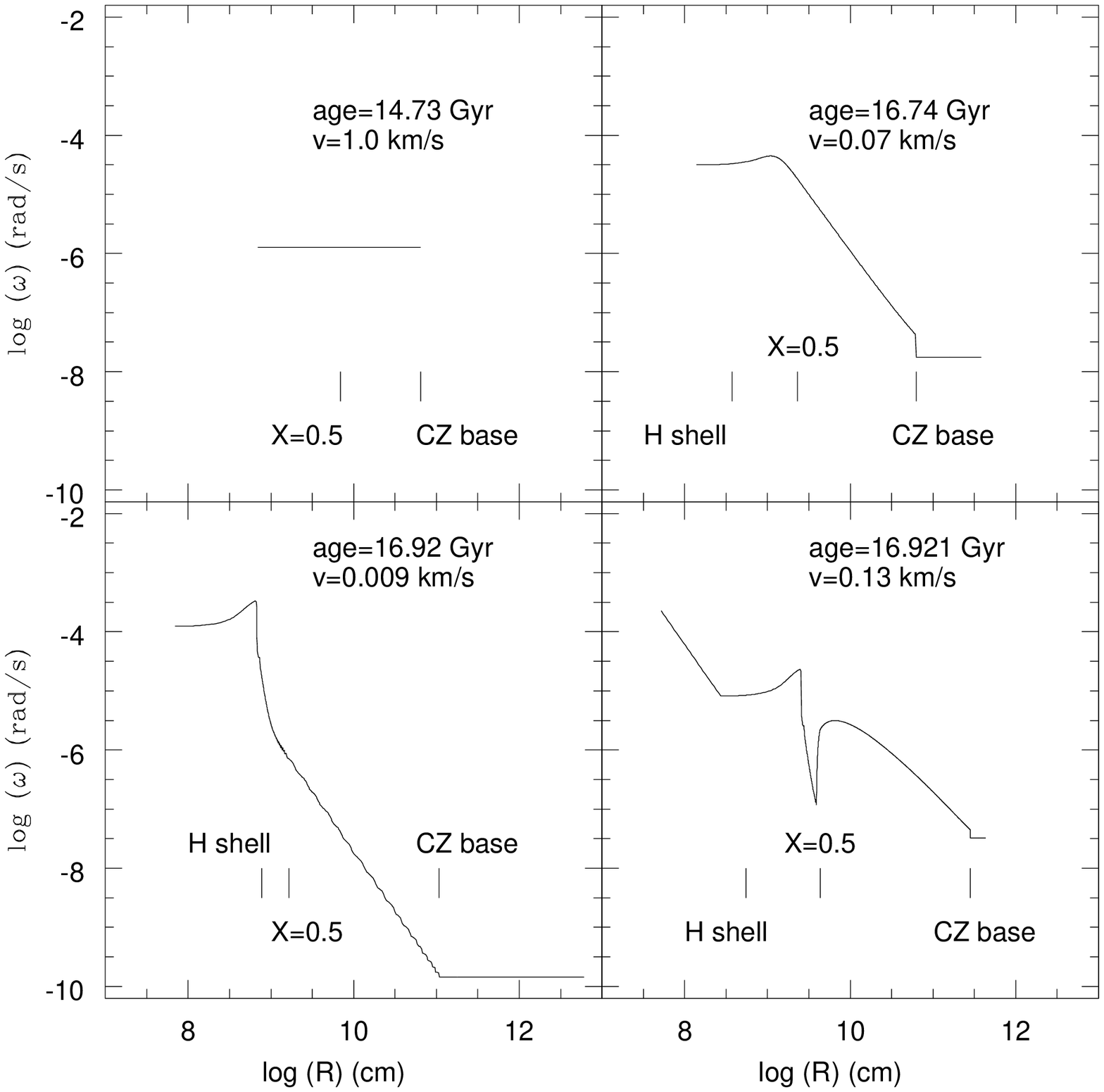}
\caption {The angular velocity evolution of a star as it evolves from
the turnoff to the horizontal branch. The $0.8 M_{\sun}$ star begins
as a solid body rotator, and then evolves assuming local conservation
of angular momentum in radiative regions, and solid body rotation in
convective regions.  There is no mass or angular momentum loss. The
four panels show the angular velocity as a function of radius for the
turnoff, the position of the maximum convection zone depth in mass,
the tip of the giant branch, and the horizontal branch. The tick marks
on each panel give the location of (from the center to the surface)
the hydrogen burning shell (not present in the turnoff model); the
position at which the hydrogen mass fraction is 50\%, and the base of
the surface convection zone. The ages and surface rotation velocities
are given in each panel.}
\end{figure}

\begin{figure}
\figurenum{2}
\plotone{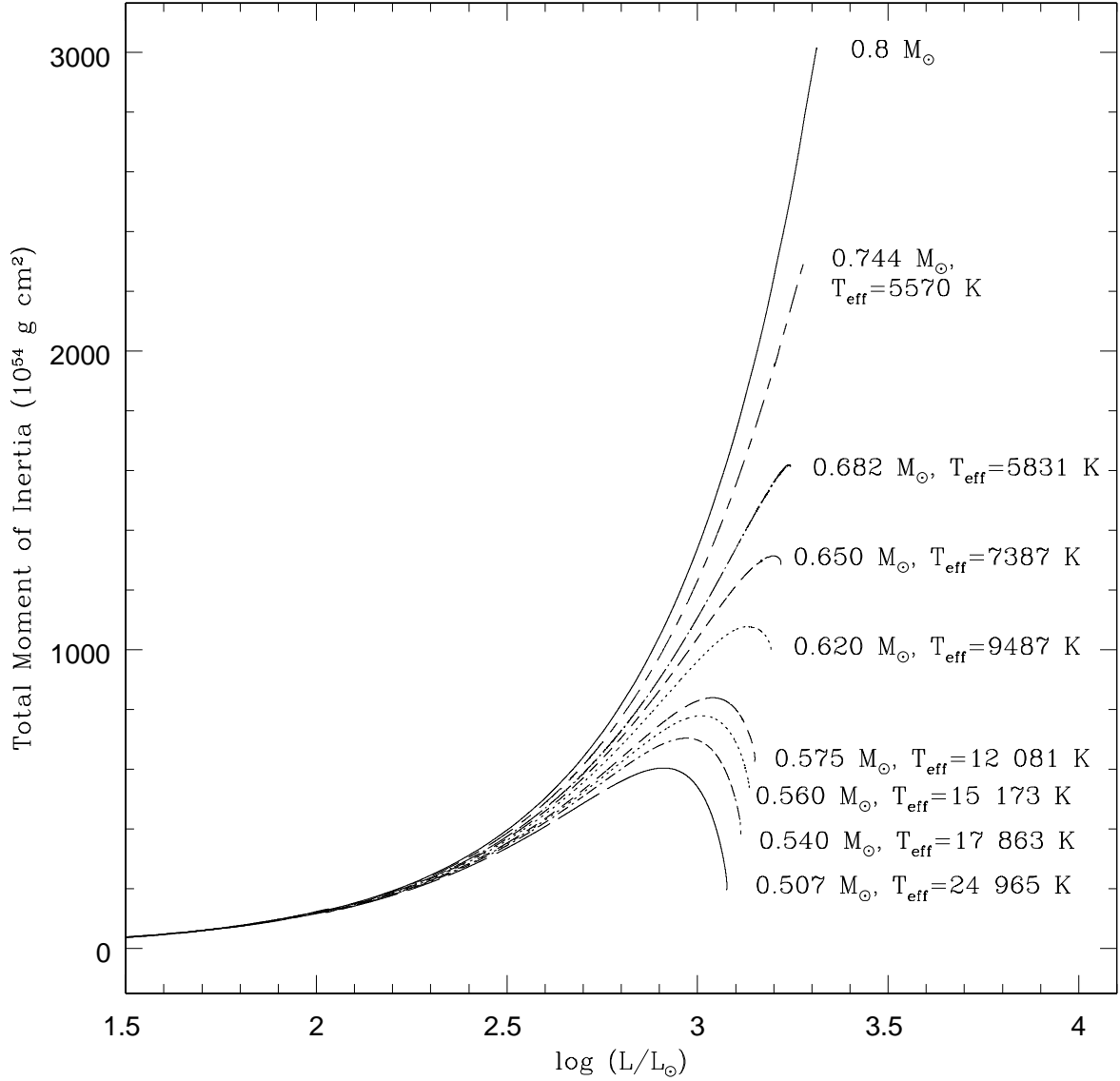}
\caption {Total moment of inertia as a function of luminosity up the
giant branch for tracks with different amount of mass loss. The final
mass of the star and its effective temperature on the horizontal
branch is given on the right side of the graph.}
\end{figure}

\begin{figure}
\figurenum{3}
\plotone{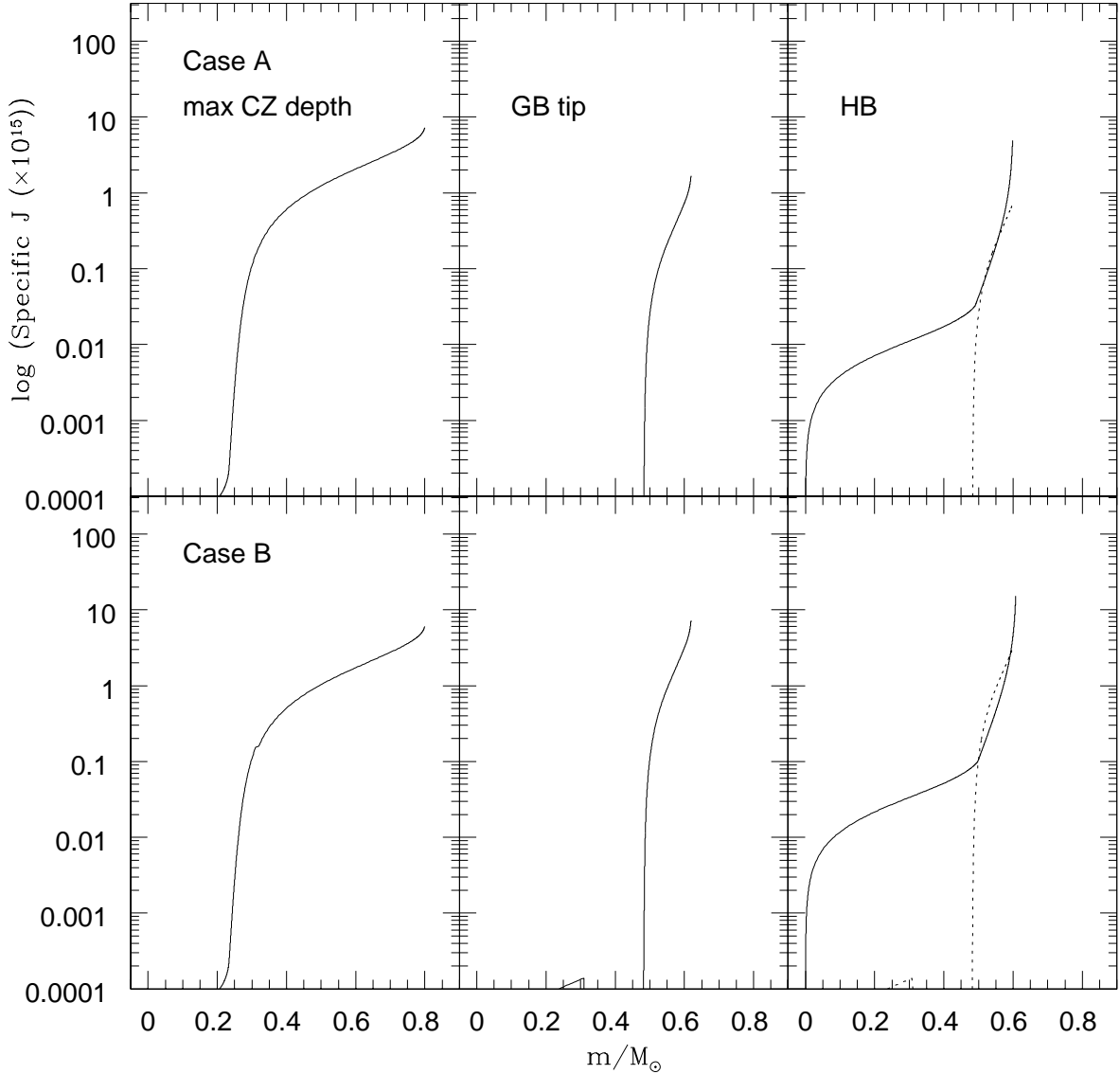}
\caption {The specific angular momentum profiles for a star which
begins its life as a $0.8 M_{\sun}$ star at the turnoff, and then
loses mass to become a $0.6 M_{\sun}$ star on the horizontal
branch. The three panels give the J/M profile for the star at the
position of the maximum convection zone depth in mass fraction, the
giant branch tip, and at horizontal branch profile. The solid line in
the third panel assumes a solid body rotation profile throughout the
star, and the dotted line shows the profile if there is local
conservation of angular momentum between the tip of the giant branch
and the horizontal branch. The top row of panels show the case in
which solid body rotation is enforced throughout the star at all times
(case A), and the lower panels show the case in which the core retains
its initial angular momentum and the convection zone rotates as a
solid body (case B).}
\end{figure}

\begin{figure}
\figurenum{4}
\plotone{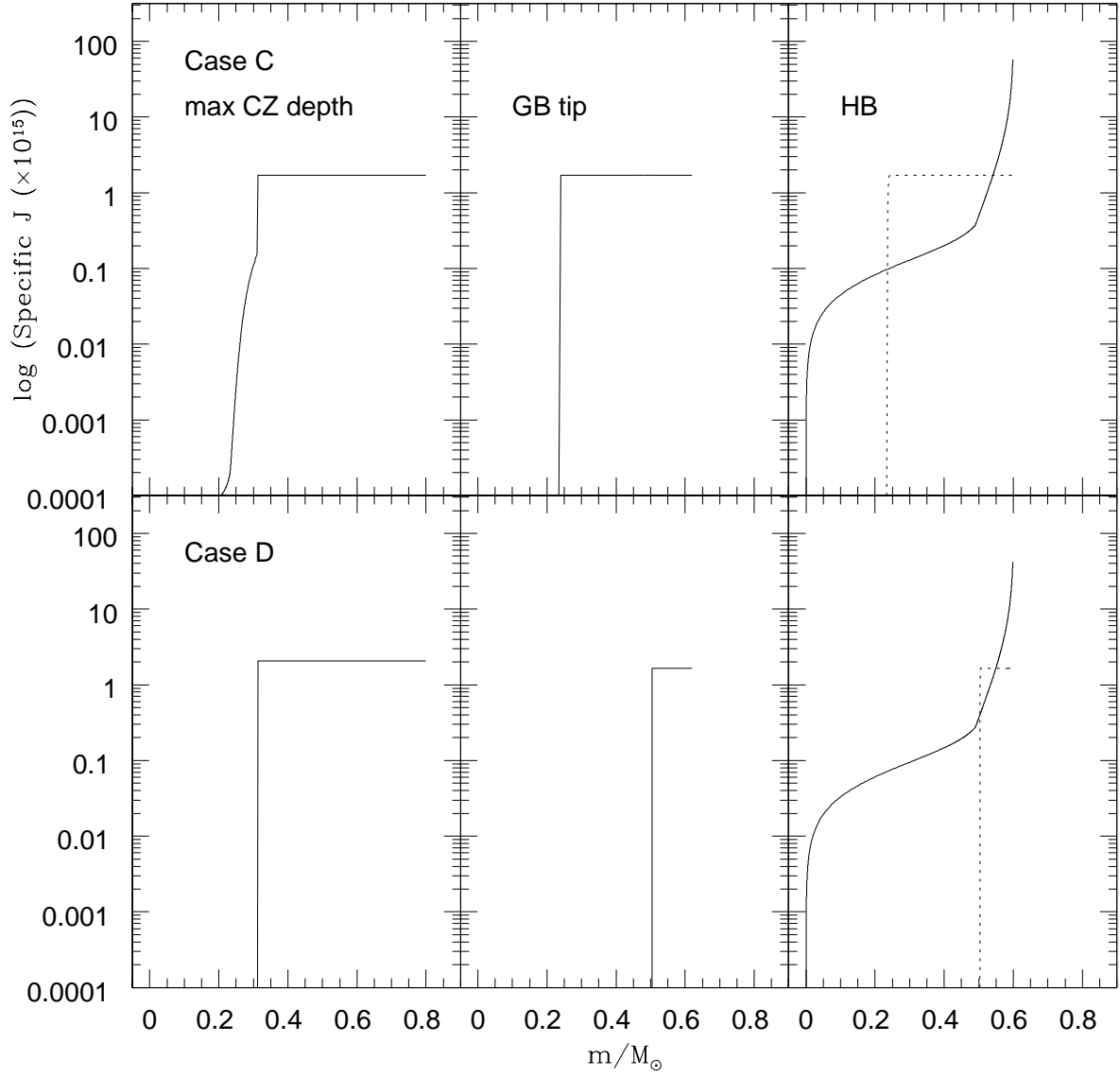}
\caption {As in figure 3, but for the two cases in which the
convection zone has constant specific angular momentum. The top panels
show the case in which the core retains the angular momentum with
which it begins at the turnoff (Case C), and the lower panels show the
case which has all the angular momentum is constrained to be in the
surface convection zone (Case D).}
\end{figure}

\begin{figure}
\figurenum{5}
\plotone{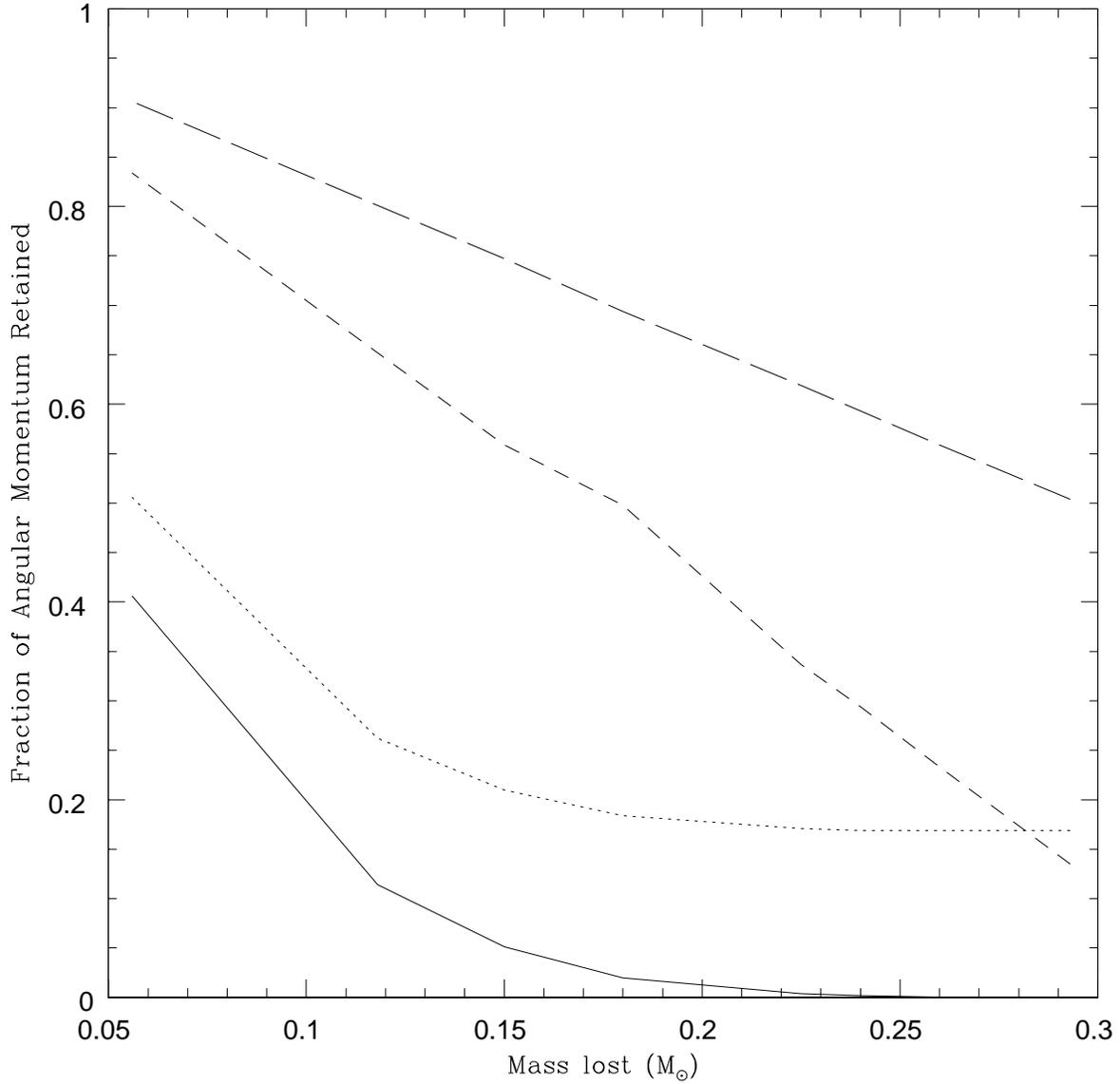}
\caption {Fraction of remaining angular momentum as a function of mass
lost for different internal angular momentum profiles. The solid line
is the case in which solid body rotation is enforced throughout the
star (case A). The dotted line shows the solid body in the convection
zone case (case B), the long dashed line has constant specific angular
momentum in the convection zone and no angular momentum in the core
(case C), and the short dashed line has constant specific angular
momentum in the convection zone with a reservoir of angular momentum
retained in the core (caseD).}
\end{figure}

\begin{figure}
\figurenum{6}
\plotone{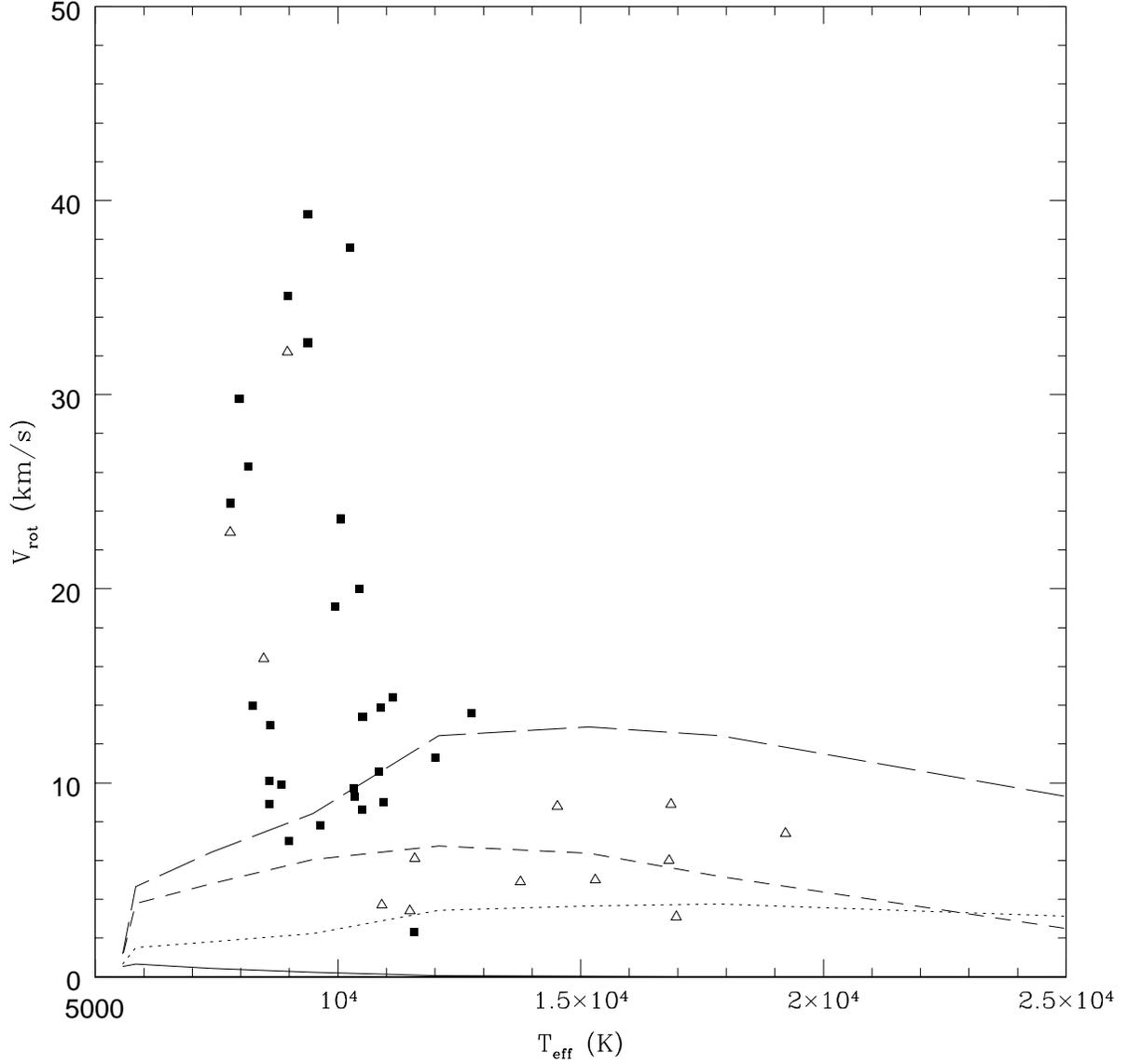}
\caption {Rotational velocity as a function of effective temperature
on the horizontal branch for different internal angular momentum
profiles. The initial angular momentum at the turnoff was assumed to
be $1 \times 10^{47}$ g cm$^2$ s$^{-1}$. The line styles correspond to
the same cases as in figure 3. The data points are taken from
\cite{PRC95} (solid squares) and \cite{behr99} (open triangles).}
\end{figure}

\begin{figure}
\figurenum{7}
\plotone{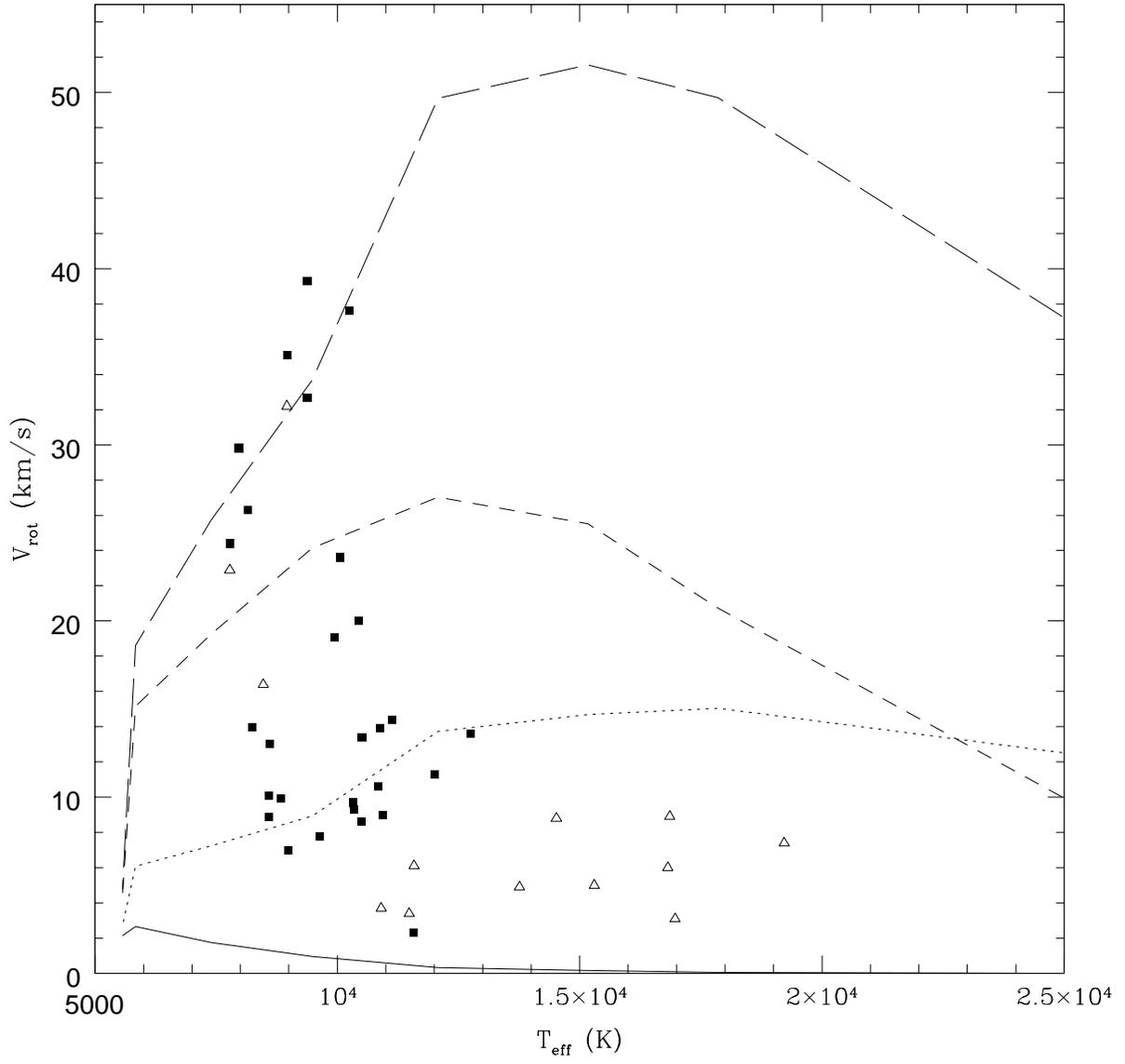}
\caption {Same as figure 4, but with an initial angular momentum
budget of $2 \times 10^{48}$ g cm$^2$ s$^{-1}$.}
\end{figure}

\clearpage
\begin{deluxetable}{ccccc}
\tablewidth{12cm}
\tablecaption{Structure of the Horizontal Branch Stars}
\tablehead{
\colhead{$\alpha$} & \colhead{M$_{HB}$ (M$_{\odot}$)} & \colhead{R$_{HB}$
(cm)} & \colhead{T$_{eff}$ (K)} & \colhead{$I_{HB}$ (g cm$^2$)} 
}
\startdata
$1.0  \times 10^{-5}$ &  0.507 & $ 1.65 \times 10^{10}$ & 24965 & $ 4.46 \times 10^{51}$ \\
$8.75 \times 10^{-6}$ &  0.540 & $ 3.51 \times 10^{10}$ & 17863 & $ 7.89 \times 10^{51}$ \\
$8.0  \times 10^{-6}$ &  0.560 & $ 5.25 \times 10^{10}$ & 15173 & $ 1.21 \times 10^{52}$ \\
$7.5  \times 10^{-6}$ &  0.575 & $ 9.75 \times 10^{10}$ & 12081 & $ 2.43 \times 10^{51}$ \\
$6.0  \times 10^{-6}$ &  0.620 & $ 1.51 \times 10^{11}$ &  9487 & $ 6.22 \times 10^{52}$ \\
$5.0  \times 10^{-6}$ &  0.650 & $ 2.65 \times 10^{11}$ &  7387 & $ 1.54 \times 10^{53}$ \\
$4.0  \times 10^{-6}$ &  0.682 & $ 4.61 \times 10^{11}$ &  5831 & $ 3.97 \times 10^{53}$ \\
$2.0  \times 10^{-6}$ &  0.744 & $ 5.11 \times 10^{11}$ &  5570 & $ 1.94 \times 10^{54}$ \\
\enddata
\end{deluxetable}

\begin{deluxetable}{ccccc}
\tablecaption{Fraction of Angular Momentum Retained}
\tablewidth{10cm}
\tablehead{
\colhead{M$_{HB}$ (M$_{\odot}$)} & \colhead {Case A} & \colhead
{Case B} & \colhead {Case C} & \colhead {Case D}
}
\startdata
0.507 & 0.00004 & 0.169 & 0.135 & 0.504 \\
0.540 & 0.0006  & 0.169 & 0.233 & 0.559 \\
0.560 & 0.002   & 0.169 & 0.294 & 0.594 \\
0.575 & 0.004   & 0.171 & 0.337 & 0.619 \\
0.620 & 0.020   & 0.184 & 0.498 & 0.694 \\
0.650 & 0.051   & 0.210 & 0.559 & 0.747 \\
0.682 & 0.114   & 0.262 & 0.652 & 0.801 \\
0.744 & 0.406   & 0.506 & 0.834 & 0.906 \\

\enddata
\tablecomments{Case A: Solid body rotation throughout the star; Case
B: convection zone rotates as a solid body and core retains its
angular momentum; Case C: constant specific angular momentum in the
convection zone and no angular momentum in the core; Case D: constant
specific angular momentum in the convection zone, the core retains its
angular momentum.}
\end{deluxetable}

\begin{deluxetable}{ccccccccc}
\tablewidth{18cm}
\tablecaption{Rotational Velocity}
\tablehead{
\colhead{M$_{HB}$ } & \multicolumn{8}{c}{Rotational Velocity
(km s$^{-1}$), assuming $J_{init} = 5 \times 10 ^{47}$ g cm$^2$
s$^{-1}$} \\
\colhead{(M$_{\odot}$)} & \colhead{Case A-1} & \colhead{Case B-1} &
\colhead{Case C-1} & \colhead{Case D-1} & \colhead{Case A-2} &
\colhead{Case B-2} & \colhead{Case C-2} & \colhead{Case D-2} 
}
\startdata
0.507 & 0.00 & 3.13 & 2.50 &  9.32 & 0.00 & 0.00 & 0.92 & 0.39 \\
0.540 & 0.01 & 3.76 & 5.18 & 12.43 & 0.01 & 0.00 & 0.40 & 0.18 \\
0.560 & 0.04 & 3.67 & 6.38 & 12.89 & 0.01 & 0.00 & 0.26 & 0.12 \\
0.575 & 0.08 & 3.43 & 6.76 & 12.42 & 0.01 & 0.00 & 0.14 & 0.07 \\
0.620 & 0.24 & 2.23 & 6.04 &  8.42 & 0.02 & 0.01 & 0.09 & 0.04 \\
0.650 & 0.44 & 1.81 & 4.81 &  6.43 & 0.02 & 0.02 & 0.50 & 0.02 \\
0.682 & 0.66 & 1.52 & 3.79 &  4.65 & 0.02 & 0.02 & 0.03 & 0.01 \\
0.744 & 0.53 & 0.67 & 1.10 &  1.19 & 0.06 & 0.01 & 0.03 & 0.01 \\
\enddata
\tablecomments{Cases A-D are the same as in figures 2 and 3, and in
table 2. The notation `1' indicates models which rotate as solid bodies
on the horizontal branch, and `2' indicates models in which local
conservation of angular momentum is imposed between the tip of the
giant branch and the horizontal branch.}
\end{deluxetable}

\end{document}